\documentclass[12pt]{article}
\usepackage[utf8]{inputenc}
\usepackage{mathptmx}
\usepackage[margin=1.25in]{geometry}
\usepackage[pagewise]{lineno}
\usepackage{xcolor}
\usepackage[utf8]{inputenc}
\usepackage{graphicx}
\usepackage{hyperref}
\usepackage{multicol}
\usepackage{amsmath}
\usepackage{amssymb}
\usepackage{lipsum}
\usepackage{booktabs}

\graphicspath{{Images/}}

\begin{document}
\begin{titlepage}
    \begin{center}
        \vspace*{1cm}
 
        \textbf{PREDICTING STATION-LEVEL BIKE-SHARING DEMANDS USING GRAPH CONVOLUTIONAL NEURAL NETWORK}
 
        \vspace{0.5cm}
 
        \vspace{1.5cm}
        \textbf{Lei Lin} \\
        NEXTRANS Center  \\
        Purdue University  \\
        E-mail: lin954@purdue.edu \\
        
        \vspace{1.5cm}
        \textbf{Weizi Li} \\
        Department of Computer Science \\
        University of North Carolina at Chapel Hill \\
        E-mail: weizili@cs.unc.edu \\
        
        \vspace{1.5cm}
        \textbf{Srinivas Peeta, Corresponding Author} \\
        Lyles School of Civil Engineering \\
        Purdue University  \\
        E-mail: peeta@purdue.edu \\
        \vspace{1.5cm}

    \end{center}
\end{titlepage}
\thispagestyle{plain}

\section*{Introduction}
A typical motorized-passenger vehicle emits about 4.7 metric tons of carbon dioxide per year~\cite{epa2005emission}. In order to decrease tailpipe emissions, reduce energy consumption, and protect the environment, as of December 2016, roughly 1000 cities around the world have started using the Bike-Sharing System (BBS)~\cite{lin2013hub}. 

While bike sharing can enhance urban mobility as a sustainable transportation mode, it has key limitations due to the effects of fluctuating spatial and temporal demands. As pointed out by many previous studies~\cite{chen2016dynamic,li2015traffic,zhou2015understanding}, it is commonly seen, for BSSs with fixed stations, that some stations are empty with no bikes to check out while others are full precluding bikes from being returned at those locations. For non-dock BSSs, enhanced flexibility poses even more challenges to ensure bike availability at some places and to prevent surplus bikes from blocking sidewalks and parking areas. For both types of BSSs, accurate bike-sharing demand predictions are critical. As a result, such a topic has attracted many research efforts~\cite{chen2016dynamic,li2015traffic,rixey2013station,faghih2014land}. 

In particular, graph convolutional neural networks (GCNNs) have been proposed to handle this problem with promising performance~\cite{shuman2013emerging,bruna2013spectral,sandryhaila2013discrete,kipf2016semi}. We propose a novel GCNN model using data-driven graph filter (GCNN-DDGF). The model does not require the predefinition of an adjacency matrix, thus can be used to learn the hidden correlations among BSS stations. Two possible architectures of the GCNN-DDGF model are developed, namely $\text{GCNN}_{\text{reg}}\text{-DDGF}$ and $\text{GCNN}_{\text{rec}}\text{-DDGF}$. The former is a regular GCNN-DDGF model which mainly consists of two types of blocks: the convolution block and the feedforward block. The latter captures temporal dependencies in bike-sharing demand series by introducing one more block---the recurrent block from the Long Short-term Memory (LSTM) neural network. To the best of our knowledge, this is the first study proposing a deep learning model for predicting station-level hourly demands by utilizing underlying correlations among stations. 

For comparison, we use four additional GCNNs, built based on a bike-sharing graph with stations as vertices. The adjacency matrices in these GCNNs are pre-defined. Together, the six GCNN models as well as seven benchmark models are evaluated using the Citi BSS dataset from New York City. Our results show that the $\text{GCNN}_{\text{rec}}\text{-DDGF}$ outperforms the rest of the models, contributing to its ability to capture hidden heterogeneous correlations among BBS stations and temporal dependencies in bike-sharing demand series. 

\section*{GCNN with Data-Driven Graph Filter}
\subsection*{Data-driven Graph Filter}
In GCNNs, the predefinition of the adjacency matrix $\tilde{A}$ is not trivial. The hidden correlations among stations may be heterogeneous. Hence, it may be hard to encode them using just one kind of metric such as the Sparse Distance (SD), Demand (DE), Average Trip Duration (ATD) or Demand Correlation (DC) matrix. Now, suppose the adjacency matrix $\tilde{A}$ is unknown; let $\hat{A} = \tilde{D}^{\frac{-1}{2}}\tilde{A}\tilde{D}^{\frac{-1}{2}}$ ($\hat{A}\in \mathbb{R}^{N \times N}$), then we have $H^l=\sigma(\hat{A}H^{l-1}W^l)$ ($H^l\in \mathbb{R}^{N \times C}$), where $\hat{A}$ is called the Data-driven Graph Filter (DDGF) and is a symmetric matrix consisting of trainable filter parameters.

The graph filter $\hat{A}$ can be learned during the training of the deep learning model. This DDGF can learn hidden heterogeneous pairwise correlations among stations to improve prediction performance. We refer such a GCNN model as GCNN-DDGF. We can view data-driven graph filtering as filtering in the vertex domain, which avoids operations such as graph Fourier transform, filtering, and inverse graph Fourier transform.

\subsection*{Architecture Design}
We explore two possible architectures of the GCNN-DDGF. The first, GCNNreg-DDGF, contains two types of blocks, the convolution block and the feedforward block. In the first step, through the convolution block, the signal vector at each station vertex is amplified or attenuated, and linearly combined with signals at other vertices weighted proportionally to the learned degrees of their correlations. The signal vectors become $(\hat{A}H^{l-1})_i$,  $(\hat{A}H^{l-1})_j$ and $(\hat{A}H^{l-1})_k$. In the second step, the signal vectors at the vertices of the next layer $l$ are calculated using the traditional feedforward block (the basic block in neural network models) to form the new signal vectors at Layer $H^l_i$, $H^l_j$, and $H^l_k$. The dimension of the vector at each vertex changes from $C^{l-1}$ to $C^l$. Suppose the $\text{GCNN}_{\text{reg}}\text{-DDGF}$ model has layers from $0,1,...$ to $m$ from the input to the output, then, the first and second steps perform the layer-wise calculation from layer $l-1$ to $l,l=1,...,m$.

The second architecture, $\text{GCNN}_{\text{rec}}\text{-DDGF}$ imports an additional block from the Long Short-term Memory (LSTM) neural network. The LSTM model is well-suited to capture temporal dependencies in time series data~\cite{hochreiter1997long}. Recently, the integration of the LSTM architecture with the CNN architecture has been reported to improve large-scale taxi demand predictions by modeling both spatial and temporal relationships~\cite{ke2017short,yao2018deep}. Hence, we expect that the introduction of the recurrent block in $\text{GCNN}_{\text{rec}}\text{-DDGF}$ can improve bike-sharing-demand prediction.

\section*{Model Development and Results}
\subsection*{Citi Bike-sharing Demand Dataset}
Our evaluation dataset contains over 28 million bike-sharing transactions between July 1st, 2013, and June 30th, 2016, from Citi BSS in New York City~\cite{citi2017}. Each transaction record includes information such as trip duration, bike check out/in time, start and end station names, start and end station latitudes/longitudes, user ID, and user type (i.e., Customer or Subscriber). 

\subsection*{Data Processing}
Suppose the bike-sharing demands for all stations in hour $i$ are $x_i \in \mathbb{R}^N$. Then, using the demand from the previous $C^0-1$ hours, we can construct a feature matrix $X_i \in \mathbb{R}^{N \times C^0}$, $X_i = [X_{i-C^0+1},...,X_i]$, and the corresponding target vector $y_{i+1} \in \mathbb{R}^N $ which represents bike-sharing demands of all stations in the next hour. The original training dataset is transferred into paired records ${(X,y)}$. The Min-Max normalization is applied to scale the data to the range of $0$ to $1$. Some previous studies regarding short-term demand forecasting have shown that utilizing latest-demand information is sufficient to generate accurate predictions~\cite{ke2017short,vlahogianni2007spatio}. 

We have built six GCNN models based on how the adjacency matrix of them is generated. These models are referred to as GCNN-SD, GCNN-DE, GCNN-ATD, GCNN-DC, $\text{GCNN}_{\text{reg}}\text{-DDGF}$, and $\text{GCNN}_{\text{rec}}\text{-DDGF}$. Their performance is evaluated using the Root Mean Square Error (RMSE) as the main criterion: $RMSE = \sqrt{\frac{1}{M*N}\sum_i^M\sum_j^N(y_{ij}-P_{ij})^2}$, where $M$ is the number of hours, $N$ is the number of stations, and $P_{ij}$ and $y_{ij}$ are the predicted and recorded bike demands in hour $i$ for station $j$, respectively.

\begin{table*}[ht] 
    \centering
    \scalebox{1}{
    \begin{tabular}{ccccc}
        \toprule  
        Model & RMSE & RMSE (7AM--9PM) & MAE & $R^2$ \\
                \midrule
        \textbf{$\text{GCNN}_{\text{rec}}\text{-DDGF}$} & \textbf{2.12} & \textbf{2.58} & \textbf{1.26} & \textbf{0.75} \\
                \midrule
        \textbf{$\text{GCNN}_{\text{reg}}\text{-DDGF}$} & \textbf{2.35} & \textbf{2.85} & \textbf{1.43} & \textbf{0.7} \\
                \midrule
        XGBoost & 2.43 & 2.95 & 1.44 & 0.68 \\
                \midrule
        LSTM & 2.46 & 3 & 1.44 & 0.67 \\
                \midrule
        GCNN-DC & 2.5 & 3.02 & 1.53 & 0.66 \\
                \midrule
        MLP & 2.51 & 3.05 & 1.51 & 0.65 \\
                \midrule
        GCNN-DE & 2.67 & 3.21 & 1.6 & 0.61 \\
                \midrule
        SVR-RBF & 2.67 & 3.25 & 1.57 & 0.61 \\
                \midrule
        LASSO & 2.7 & 3.27 & 1.65 & 0.6 \\
                \midrule
        SVR-linear & 2.72 & 3.31 & 1.52 & 0.59 \\
                \midrule
        GCNN-SD & 2.77 & 3.31 & 1.68 & 0.58 \\
                \midrule
        HA & 3.44 & 3.42 & 2.08 & 0.35 \\
                \midrule
        GCNN-ATD & 3.44 & 3.83 & 2.21 & 0.35 \\
        \bottomrule
    \end{tabular}}
    \caption{Comparison in model performance using the test dataset.}
    \label{tb:results}
\end{table*}

We show the model performance in Table~\ref{tb:results}. In addition to RMSE, Mean Absolute Error (MAE) and $\text{R}^2$ are used for evaluation. We calculate RMSE over the period 7AM to 9PM, since bike-sharing demands over other time periods are mostly zero or close to zero. As a result, $\text{GCNN}_{\text{rec}}\text{-DDGF}$ performs the best under all measures. It has the lowest RMSE (2.12), RMSE (7AM--9PM) (2.58), and MAE (1.26), and the highest $\text{R}^2$ (0.75). $\text{GCNN}_{\text{reg}}\text{-DDGF}$ performs the second best, which indicates that the design of DDGF and the usage of the recurrent block from LSTM are effective. 

The performance of the two GCNN-DDGF models are followed by XGBoost and LSTM. While XGBoost is not designed to capture temporal dependencies in the bike-sharing demand series or the hidden correlations among stations, it supports fine-tuning and regularization for preventing overfitting~\cite{zhu2017prediction}. LSTM performs closely to XGBoost by utilizing temporal dependencies in the bike-sharing demand series. The next best performance is from GCNN-DC, in which the pre-defined adjacency matrix with the Pearson Correlation Coefficient makes it the best among the four GCNNs with pre-defined adjacency matrices. The GCNN-ATD model performs the worst, and has the largest RMSE (3.44), RMSE (7AM--9PM) (3.83), and MAE (2.21), and the lowest $\text{R}^2$ (0.35). This indicates that ATD is not suitable for a graph adjacency matrix. It also shows that the quality of the adjacency matrix has a huge impact on the performance of the GCNN model. The remaining benchmark models perform poorly as they do not factor correlations among stations or temporal dependencies in time series.

\section*{Conclusion and Future Research Directions}
We have proposed a novel GCNN-DDGF model for station-level hourly demand prediction in a large-scale bike-sharing network. Different from the state-of-the-art CNN model, the GCNN model does not require data to have a regular grid structure. Consequently, it can be used to address many graph-based problems including transportation-related applications. We have implemented four GCNN models with adjacency matrices from multiple BSS data such as the SD, DE, ATD, and DC matrices. Furthermore, we have explored two architectures: $\text{GCNN}_{\text{reg}}\text{-DDGF}$ and $\text{GCNN}_{\text{rec}}\text{-DDGF}$. Both models can address the limitations of GCNN, which performance relies on a pre-defined graph structure. $\text{GCNN}_{\text{rec}}\text{-DDGF}$ also implements the Long Short-term Memory (LSTM) neural network for capturing the temporal dependencies in bike-sharing demand series.

The six GCNN models and seven other benchmark models are built and evaluated using the Citi BSS dataset from New York City, which includes over 28 million transactions from 2013 to 2016. RMSE, MAE, and $\text{R}^2$ are used as measuring criteria. Our results show that $\text{GCNN}_{\text{rec}}\text{-DDGF}$ performs the best under all measurements, followed by $\text{GCNN}_{\text{rec}}\text{-DDGF}$.  GCNN-ATD performs the worst. This observation confirms the insight from previous studies, which states that the performance of GCNN depends heavily on the pre-defined structure of the graph. 

In future research, first, we would like to consider more factors such as weather and social events (holidays and sports games). These variables can be concatenated with the input layer of the feedforward block of GCNN-DDGF. Second, the current model can be modified to be an online, real-time algorithm in order to process mobile traffic data~\cite{Lin2019ComSense}. Third, we would like to test our model on other transportation problems such as subway station demand prediction, and network-wide traffic state estimation and reconstruction~\cite{Li2017CityEstSparse,Li2017CityFlowRecon,Li2018CityEstIter}. Fourth, it would be useful to derive a model that can learn a sparse graph filter capturing directional relationships among bike-sharing stations. Finally, we are interested in using GCNN-DDGF to enhance the heterogeneity and accuracy of traffic simulation models~\cite{Wilkie2015Virtual,Chao2019Survey} and to study the interplay between the bike-sharing system and connected and autonomous vehicles in a city~\cite{Li2019ADAPS}.

\bibliographystyle{ieeetr}
\bibliography{references}

\end{document}